\documentclass[10pt,conference]{IEEEtran}
\IEEEoverridecommandlockouts
\usepackage{cite}
\usepackage{amsmath,amssymb,amsfonts}
\usepackage{algorithmic}
\usepackage{graphicx}
\usepackage{textcomp}
\usepackage{xcolor}
\usepackage{orcidlink}
\def\BibTeX{{\rm B\kern-.05em{\sc i\kern-.025em b}\kern-.08em
T\kern-.1667em\lower.7ex\hbox{E}\kern-.125emX}}

\usepackage{blindtext}
\usepackage{verbatim}
\usepackage{url}
\usepackage{svg}
\usepackage{multirow}
\usepackage{subcaption}
\usepackage{graphicx}
\usepackage{tikz}
\usepackage{nicefrac}
\usetikzlibrary{quantikz2}
\usetikzlibrary{shapes.geometric}
\tikzset{slice/.append style={line width=1.5pt}}

\usepackage{standalone}
\usepackage{tikzit}

\usepackage{wasysym}
\usepackage{textcomp}
\usepackage{makecell}
\usepackage{physics}

\usepackage{hyperref}
\hypersetup{
    colorlinks,
    linkcolor={red!80!black},
    citecolor={green!80!black},
    urlcolor={blue!80!black}
}

\usepackage{cleveref}
\Crefname{section}{Sec.}{Secs.}
\Crefname{theorem}{Thm.}{Thms.}
\Crefname{lemma}{Lem.}{Lems.}
\Crefname{corollary}{Cor.}{Cors.}
\Crefname{definition}{Def.}{Defs.}
\Crefname{remark}{Rem.}{Rems.}
\Crefname{figure}{Fig.}{Figs.}
\Crefname{equation}{Eq.}{Eqs.}

\begin{document}

\title{CUAOA: A Novel CUDA-Accelerated\\Simulation Framework for the QAOA}


\makeatletter
\newcommand{\linebreakand}{%
  \end{@IEEEauthorhalign}
  \hfill\mbox{}\par
  \mbox{}\hfill\begin{@IEEEauthorhalign}
}
\makeatother

\author{
\IEEEauthorblockN{Jonas Stein\textsuperscript{$\orcidlink{0000-0001-5727-9151}$}}
\IEEEauthorblockA{\textit{LMU Munich, Germany}\\\textit{Aqarios GmbH, Germany}\\jonas.stein@ifi.lmu.de}
\and
\IEEEauthorblockN{Jonas Blenninger\textsuperscript{$\orcidlink{0009-0004-5382-7113}$}}
\IEEEauthorblockA{\textit{LMU Munich, Germany}\\\textit{Aqarios GmbH, Germany}\\jonas.blenninger@aqarios.com}
\and
\IEEEauthorblockN{David Bucher\textsuperscript{$\orcidlink{0009-0002-0764-9606}$}}
\IEEEauthorblockA{\textit{Aqarios GmbH, Germany}\\david.bucher@aqarios.com}
\linebreakand 
\IEEEauthorblockN{Peter J. Eder\textsuperscript{$\orcidlink{0009-0006-3244-875X}$}}
\IEEEauthorblockA{\textit{Siemens AG, Munich, Germany}\\peter-josef.eder@siemens.com}
\and
\IEEEauthorblockN{Elif Çetiner}
\IEEEauthorblockA{\textit{LMU Munich, Germany}\\elif.cetiner@tum.de}
\and
\IEEEauthorblockN{Maximilian Zorn\textsuperscript{$\orcidlink{0009-0006-2750-7495}$}}
\IEEEauthorblockA{\textit{LMU Munich, Germany}\\maximilian.zorn@ifi.lmu.de}
\and
\IEEEauthorblockN{Claudia Linnhoff-Popien\textsuperscript{$\orcidlink{0000-0001-6284-9286}$}}
\IEEEauthorblockA{\textit{LMU Munich, Germany }\\linnhoff@ifi.lmu.de}
}

\maketitle

\bstctlcite{BSTcontrol}

\begin{abstract}
The Quantum Approximate Optimization Algorithm (QAOA) is a prominent quantum algorithm designed to find approximate solutions to combinatorial optimization problems, which are challenging for classical computers. In the current era, where quantum hardware is constrained by noise and limited qubit availability, simulating the QAOA remains essential for research. However, existing state-of-the-art simulation frameworks suffer from long execution times or lack comprehensive functionality, usability, and versatility, often requiring users to implement essential features themselves. Additionally, these frameworks are primarily restricted to Python, limiting their use in safer and faster languages like Rust, which offer, e.g., advanced parallelization capabilities. In this paper, we develop a GPU accelerated QAOA simulation framework utilizing the NVIDIA CUDA toolkit. This framework offers a complete interface for QAOA simulations, enabling the calculation of (exact) expectation values, direct access to the statevector, fast sampling, and high-performance optimization methods using an advanced state-of-the-art gradient calculation technique. The framework is designed for use in Python and Rust, providing flexibility for integration into a wide range of applications, including those requiring fast algorithm implementations leveraging QAOA at its core. The new framework's performance is rigorously benchmarked on the MaxCut problem and compared against the current state-of-the-art general-purpose quantum circuit simulation frameworks Qiskit and Pennylane as well as the specialized QAOA simulation tool QOKit. Our evaluation shows that our approach outperforms the existing state-of-the-art solutions in terms of runtime up to multiple orders of magnitude. Our implementation is publicly available at \url{https://github.com/JFLXB/cuaoa} and Zenodo~\cite{blen_CUAOA_2024}.
\end{abstract}


\section{Introduction}
\label{sec:introduction}
One of the most promising approaches for quantum advantage in the domain of optimization problems is the Quantum Approximate Optimization Algorithm (QAOA)~\cite{farhi2014quantum}. As a parameterized version of the Quantum Adiabatic Algorithm~\cite{farhi2000quantum}, the QAOA utilizes a quantum phenomenon, i.e., the Adiabatic Theorem~\cite{Born1928}, to solve optimization problems approximatively. In consequence of many results about improved scaling performance of the QAOA for important optimization problems compared to classical state-of-the-art solvers \cite{boulebnane2022solving,doi:10.1126/sciadv.adm6761,carlson2023quantumadvantageclassicallocal}, much scientific effort goes into achieving large scale numerical simulations to explore possible quantum advantages empirically~\cite{PhysRevResearch.6.013223,Medvidovic2021,Shang2023}. 

The current standard for large-scale, noise-free quantum circuit simulations relies on efficient matrix multiplication via GPUs and more specifically the cuStateVec SDK~\cite{10313722}, which allows for a translation of circuit instructions written in Qiskit~\cite{qiskit2024}, Pennylane\cite{bergholm2022pennylaneautomaticdifferentiationhybrid}, or other SDKs, to CUDA, the toolkit for instructing computations on NVIDIA GPUs~\cite{CUDA}. 

Due to its specific quantum circuit structure, the runtime complexity of simulating a standard $p$-depth, $n$-qubit, $X$-mixer-based, QAOA circuit is $\mathcal{O}(pn2^n)$ instead of the $\mathcal{O}(p4^n)$ for general $p$-depth quantum circuits~\cite{qokit}. As we expect no further speedup for the simulation of the $X$-mixer (cf. \cite{qokit,1674569}), and to allow for arbitrary mixers (cf. \cite{9259965}), we focus on speedup gains based on the diagonal structure of the cost operator (cf.~\cite{qokit}).

In the current state-of-the-art QAOA-simulator QOKit \cite{qokit}, Lykov et al. exploit this diagonal structure of the cost operator by precomputing the costs for all possible solutions in parallel and then using these for the cost unitary application as well as the expectation value calculation. However, QOKit is written in Python and hence cannot use CUDA natively, hindering itself from achieving the fastest possible computation. What is more, the two key features besides the precomputation of the cost Hamiltonian are not executable in its provided code due to missing implementation: the cuStateVec simulator and the gradient computation~\cite{LykovGithub2023}.

Resolving the shortcomings of QOKit, we develop a CUDA-based QAOA simulator (CUAOA) that is optimized to execute all practically relevant operations for QAOA-evaluation in native CUDA, oriented towards a single-GPU setting. To ensure compatibility with the expected end-user programming language Python (e.g., via PyO3~\cite{PyO3_Project_and_Contributors_PyO3}) while allowing for direct CUDA access, we employ Rust with integrated C/C++ modules (via Foreign Function Interface (FFI)). Analog to the current state of the art, we use cuStateVec for all computations that do not concern the cost unitary, e.g., the mixer application and the sampling process. Allowing for significant speedups compared to current state-of-the-art QAOA simulators, we propose CUDA-native implementations of
\begin{itemize}
    \item the precomputation of the cost Hamiltonian,
    \item the application of the cost unitary,
    \item the calculation of the expectation value, and
    \item the gradient computation.
\end{itemize}


The remainder of this paper is structured as follows. In ~\Cref{sec:background}, we outline preliminaries on the simulation of the QAOA, the adjoint differentiation method, and CUDA. In \Cref{sec:methodology}, CUAOA is presented and subsequently evaluated in \Cref{sec:evaluation}. Finally, we conclude our findings in \Cref{sec:conclusion}.

\section{Background}
\label{sec:background}
In this section, we provide preliminaries on the QAOA, a method to efficiently compute gradients in classically simulated quantum circuits and CUDA.
\subsection{The Quantum Approximate Optimization Algorithm}
Given a combinatorial optimization problem by an objective function $f:\lbrace 0,1\rbrace^n\rightarrow\mathbb{R}$, the QAOA conducts the following steps to approximate the optimal solution~\cite{farhi2014quantum}:
\begin{enumerate}
    \item Mapping the objective values onto the eigenvalues of the diagonal cost Hamiltonian $H_C=\sum_x f(x)\ket{x}\!\bra{x}$.
    \item Preparing a system in the ground state of the mixer Hamiltonian, i.e., usually $\ket{+}^{\otimes n}$ for $H_M=-\sum_{i=1}^n \sigma_i^x$.
    \item Simulating the time evolution $\exp(i\int_0^T H_s(t)\textnormal{d}t)$ approximatively, where $H_s(t)=\left(1 - s(t)\right)H_M + s(t)H_C$ governs the adiabatic evolution and a bijective $s:\left[0,T\right]\rightarrow\left[0,1\right]$ increases monotonically for given $T>0$.
    \item Measuring the resulting state and remapping it to its corresponding solution of the objective function $f$.
\end{enumerate}
To simulate the time evolution of $H_s$ in a quantum circuit, a discretization into $p\in\mathbb{N}$ Hamiltonians $H_s(1/T),...,H_s(T)$ via Trotterization is carried out, amounting the following unitary operation forming the QAOA:
\begin{align}
    U\left(\beta, \gamma\right) = U_M(\beta_p) U_C(\gamma_p) \ldots U_M(\beta_1) U_C(\gamma_1)H^{\otimes n},
\end{align}
where $\beta_i$ and $\gamma_i$ control the speed of the time evolution and $U_M(\beta_i) = e^{-i\beta_i H_{M}}$, $U_C(\gamma_i) = e^{-i\gamma_i H_{C}}$, s.t. $U\left(\beta, \gamma\right)$ approaches adiabatic time evolution for $p\rightarrow \infty$, and constant speed, i.e., $\beta_i = 1-i/p$, and $\gamma_i = i/p$~\cite{sack2021}.

\subsection{Adjoint Differentiation} \label{subsec:adjointdiff}
The adjoint differentiation method \cite{jones2020efficientcalculationgradientsclassical} exploits the possibility to clone statevectors in classical quantum circuit simulators to yield a runtime of $\mathcal{O}(P)$ instead of the generally employed parameter shift rule which has complexity $\mathcal{O}(P\cdot m)$ \cite{PhysRevA.98.032309}, where $P$ is the number of possibly parameterized layers in the quantum circuit and $m$ is the number of parameters. These complexities state the query complexity of simulating the application of a layer of gates, which is generally $\mathcal{O}(4^{n})$ for an $n$-qubit quantum register.

In the following we assume that each circuit layer $U_i$ has exactly one parameter $\theta_i$, which is the case for the standard form of QAOA. For details on adaptations necessary for this approach to work for more general circuit layers involving repeated parameters and multiple parameters per layer see Ref.~\cite[Appendix B]{jones2020efficientcalculationgradientsclassical}. Notably, these increase the runtime complexity by constant factors.

The adjoint differentiation method exploits the hermiticy of the partial derivative of the measurement operator $M$, i.e., 
\begin{align*}
    \frac{\partial\expval{M}}{\partial\theta_i}=&\bra{0}U_1^\dagger \ldots \frac{\partial U_i^\dagger}{\partial\theta_i}\ldots U_P^\dagger M U_P \ldots U_i \ldots U_1 \ket{0}\\
    & + \bra{0}U_1^\dagger \ldots U_i^\dagger \ldots U_P^\dagger M U_P \ldots \frac{\partial U_i}{\partial\theta_i} \ldots U_1 \ket{0} \\
    = & 2 \, \mathfrak{R}\left(\bra{0}U_1^\dagger \ldots U_i^\dagger \ldots U_P^\dagger M U_P \ldots \frac{\partial U_i}{\partial\theta_i} \ldots U_1 \ket{0}\right),
\end{align*}
which can be written as $\nabla_{\theta_i}\expval{M}= 2 \, \mathfrak{R}\left(\mel{b_i}{\nicefrac{\partial U_i}{\partial \theta_i}}{k_i}\right)$, where $\bra{b_i}\coloneqq \bra{0}U_1^\dagger \ldots U^\dagger_i\ldots U_P^\dagger M U_P \ldots U_{i+1}$ and $\ket{k_i}\coloneqq U_{i-1}\ldots U_1 \ket{0}$ can be computed recursively via $\bra{b_{i+1}}=\bra{b_{i}}U_{i+1}^\dagger $ and $\ket{k_{i+1}}=U_{i}\ket{k_i}$. Due to this recursive nature, it takes $\mathcal{O}(P)$ layer executions to calculate $\nabla_{\theta_1}\expval{M}$ and then $\mathcal{O}(1)$ layer executions for all other partial derivatives yielding the stated overall runtime complexity of $\mathcal{O}(P)$.

\subsection{CUDA}
The Compute Unified Device Architecture (CUDA) is a parallel computing platform
and application programming interface (API) model that enables developers to directly access NVIDIA Graphics Processing Units
(GPUs)~\cite{CUDA}. The main advantage of GPUs over CPUs lies in the ability to execute computations in a massively parallelized manner. This is especially relevant for applications like matrix multiplication, which can be divided into many smaller, independent computations.

CUDA is an extension of the C/C++ programming languages that has a hierarchical structure centered around threads. Calls from the CPU to CUDA are done via kernel functions, known as kernels, that run on the GPU. These kernels are executed by a grid of thread blocks, with each block containing multiple threads. This hierarchical arrangement enables efficient use of the GPU’s resources and provides precise control over each thread's behavior. Threads within the same block can share data through shared memory, which is unique to each block and generally faster than global memory. Communication between threads of different blocks occurs through global memory. \cite{cuda21}

\section{Related Work}
\label{sec:related-work}
While multiple frameworks exist that are aimed at providing QAOA circuit implementations of various versions of the QAOA (e.g., OpenQAOA~\cite{sharma2022openqaoasdkqaoa} and JuliaQAOA~\cite{Golden_2023}), only QOKit~\cite{qokit} is aimed at---and capable of---achieving a significant speedup through GPU usage. Therefore we focus on QOKit in the remainder of this section.

QOKit is targeted towards simulating the QAOA involving large amounts of qubits and offers a multi-GPU approach that shares information about the statevector via OpenMPI~\cite{10.1007/11752578_29}. The two key features of QOKit are a parallelized computation and application of the cost operator and an algorithm to apply the $X$-mixer in time $\mathcal{O}(n2^n)$.

At the time of this paper being published, QOKit is limited by significant shortcomings in their published code, i.e., missing support of the cuStateVec circuit simulator as well as missing gradient calculation, which both are key components for achieving the shortest runtimes possible. Furthermore, QOKit is limited by its implementation being carried out in Python, which manifests in that their proposed mixer unitary, as well as their OpenMPI-based parallelization perform worse compared to plain cuStateVec~\cite{qokit}. Also, for extracting information about the statevector, e.g., for sampling, the probabilities of the statevector are copied to the CPU, which displays a significant runtime bottleneck.

Based on the results presented in Ref.~\cite{qokit}, the only clear speedup that QOKit provides over a plain cuStateVec-based implementation appears to be the efficient precomputation of the cost operator and its application in diagonal form. The precomputation is based on the polynomial representation of the objective function
\begin{align}
    f(s)=\sum_{k=1}^L w_k \prod_{i\in t_k}s_i,
\end{align}
where $s\in\left\lbrace -1,1\right\rbrace^n$ and $\mathcal{T}\coloneqq \left\lbrace(w_1,t_1),...,(w_L,t_L)\right\rbrace$ defines the polynomial terms through the indices of the involved variables $t_k\subseteq\left\lbrace i\mid 1\leq i \leq n\right\rbrace$ and their associated weight $w_k\in\mathbb{R}$. For the computation of $f(s)$ of all possible inputs $s$, an array of zeros is allocated on the GPU and then a GPU kernel iterating over all terms in $\mathcal{T}$ is applied in parallel for each entry in the array. The value of each term is calculated using bitwise-XOR and population count operations to determine the sign of $\prod_{i\in t_k}s_i$. The application of the cost unitary is executed by an element wise product of the statevector with $\exp(-i\gamma_i f(s))$.

\section{Methodology}
\label{sec:methodology}
Aiming to exploit the diagonal structure of the cost operator in the QAOA, we now propose CUAOA, a CUDA-accelerated, single-GPU quantum circuit simulator for the QAOA that annihilates the shortcomings of QOKit. To offer the same convenience to end-users as QOKit---a Python interface---while enabling seamless CUDA-operability, the core module of CUAOA is written in Rust. This allows access to CUDA-instructions written in C/C++ via FFI and the integration from within Python via libraries like PyO3~\cite{PyO3_Project_and_Contributors_PyO3}. This also yields the significant advantage of running the exact same program instructions much faster compared to a Python implementation, just because Rust is a compiled language, i.e., the code is compiled directly to machine code before execution.

Starting from a baseline of the state of the art in circuit-agnostic noise-free quantum circuit simulation, i.e., cuStateVec, for our QAOA simulation, we now show how every circuit simulation component that involves a cost operator can be implemented more efficiently via CUDA-based implementation of the operator in its diagonal form.

\subsection{Cost function representation}
In standard QAOA simulation, the cost operator is represented by quantum gates modelling the cost function. However, as the cost operator is diagonal in most practical applications (which is a result of basis encoding), the application of the cost operator in a classical circuit simulation can be carried out directly through a multiplication with a diagonal matrix, which can be fully parallelized on GPUs.

Improving on the cost function computation of QOKit~\cite{qokit}, we work with a $0-1$ based function representation that reduces the number of necessary additions significantly: Encode the indices of the $n$ binary variables using one-hot encoding, we can represent a polynomial objective function as
\begin{align}\label{eq:costcalc}
    f(x)=\sum_{k=1}^L w_k \left(x \Leftrightarrow x \land_{\textnormal{b}} \left[\sum_{v_i\in t_k}2^i\right]_2\right),
\end{align}
where each term $t_k$ consists of variable indices $v_i\in\left[n\right]$ and $\Leftrightarrow$ denotes the logical equivalency that yields $1$ or $0$ respectively. As the binary string $\left[\sum_{v_i\in t_k}2^i\right]_2$ can easily be computed through bitwise logical AND operations (denoted $\land_{\textnormal{b}}$) of the one-hot representation of each term $t_k$, and as only the entries for which the bitstring $x$ is nonzero have to be considered for any given $x$, a large amount of terms can be ignored in computation. This shortcut is not exploited in QOKit, which iterates over all terms. Thus, our approach can be significantly faster for polynomials that have a large amount of small degree terms, which even are the norm in practice. For problems inheriting symmetries (e.g., the MaxCut problem), many cost values equal each other, such that additional speedups could be gained. However, we refrain from such optimizations to maintain problem agnosticity.

\subsection{QAOA circuit simulation}
The CUAOA starts by allocating memory for the statevector as well as the cost Hamiltonian and stores the pointers referencing the respective GPU memory. In addition, a CUDA stream is created and its reference is stored in the handle, which allows multiple kernels associated with different streams to be executed in parallel on the same GPU. Further, the handle for interactions with the cuStateVec library is initialized with the handle's stream and subsequently stored. Memory for other variables is not allocated upon the handle's initialization, but only later when they are actually needed, to reduce memory usage.

The statevector is initialized as an array of the CUDA double type for complex numbers \texttt{cuDoubleComplex} in parallel for each entry. As all evaluations are carried out for the standard form of QAOA with an $X$-Mixer, we directly initialize all values of the array to $1/\sqrt{2^n}$. The cost Hamiltonian is initialized as an array of double precision entries and the value of each entry is computed via \Cref{eq:costcalc} in parallel. The cost unitary is applied to the statevector $\ket{\psi}$ exploiting Eulers formula $\exp(i\theta) = \cos{\theta} + i\sin{\theta}$ through $\psi_i \mapsto \left[\cos(-\gamma_i f(x)) + i \sin(-\gamma_i f(x))\right] \cdot \psi_i$. The variational parameter $\gamma_i$ for this is passed as a double-precision input to this operation and the CUDA built-in function for multiplication \texttt{cuCmul} is used. Note that while cuStateVec also offers a function to directly apply a diagonal matrix\footnote{Namely \texttt{custatevecApplyGeneralizedPermutationMatrix}.}, they do not support the in-place multiplication of the $\gamma_i$, which would have to be done through another kernel leading to resource inefficiency and thus giving the stated approach its right for existence.

To apply the mixer unitary, \texttt{custatevecApplyMatrix} is used. As we only used the $X$-mixer in our evaluation, this reduces to the application of an $R_x(-2\beta_i)$ gate for every qubit.

\subsection{Gradient computation}
To compute the gradient of a QAOA circuit with respect to its variational parameters $\gamma$ and $\beta$, we employ the adjoint differentiation technique outlined in \Cref{subsec:adjointdiff}, as it is the state of the art for gradient calculation in classical circuit simulation. For the QAOA, another simplification in gradient calculation of each layer arises from the well-known identity $\frac{\partial}{\partial t}e^{tA}=Ae^{tA}$, which implies that $\frac{\partial}{\partial t}e^{-i\gamma_iH_C}=-iH_Ce^{-i\gamma_iH_C}$ and $\frac{\partial}{\partial t}e^{-i\gamma_iH_M}=-iH_Me^{-i\gamma_iH_M}$. While the application of $H_C$ is trivial, the application of $H_M$ reduces to a layer of $X$-gates for all of our evaluation runs, as we only consider the standard $X$-mixer. The uncomputation needed for each gradient calculation step described in \Cref{subsec:adjointdiff} also simplifies based on $\frac{\partial}{\partial t}e^{tA}=Ae^{tA}$, as only $-iH_M$ and $-iH_C$ respectively have to be uncomputed. As both operators are Hermitian (i.e., $H_M=H^\dagger_M$ and $H_C=H^\dagger_C$), this uncomputation can be done by applying $iH_M$ and $iH_C$ respectively. In our implementation we get rid of the introduced imaginary number $i$ by switching from the real to the imaginary part (cf. \Cref{subsec:adjointdiff}).

\subsection{Retrieving Results from the GPU}\label{subsec:expvalalgo}
Arguably the most important output of a QAOA simulator is the expectation value $\mel{\psi}{H_C}{\psi}=\sum_{i=1}^{2^n} f(x_i)\left|\psi_i\right|^2$. To compute this sum, we calculate $f(x_i)\left|\psi_i\right|^2$ for all $i$ based on the resulting QAOA statevector $\ket{\psi}$ and the cost operator, and store the result in a new array of doubles. Then we calculate the sum of all components of this array by braking it down into a tree-like hierarchical structure where, at each level, always two elements are added in parallel, amounting to a total of $\log_2(2^n)$ sequential computations.

To sample from the statevector, we use the sampling functionality offered in cuStateVec. This has the big advantage that the statevector does not have to be copied to the CPU thus evading any memory-transformation bottlenecks. In addition to the sampled solution bitstring, we also output the respective objective value, as it is stored on the GPU anyway and thus save additional computational efforts for the user. What is more, our implementation also supports the edge-case of exporting the complete statevector from GPU to CPU.

\section{Evaluation}
\label{sec:evaluation}
To evaluate the performance of CUAOA, we examine its runtime for the full circuit execution with regards to outputting the expected value as well as sampling, and also its performance in parameter training using gradient-based methods.

In alignment with QOKit's evaluation~\cite{qokit}, we consider the MaxCut problem with three types of graphs ranging from $6$ to $29$ vertices: (1) random graphs generated based off the {Erd\H{o}s-R\'{e}nyi} $G(n,p)$ model~\cite{erdos1959} with 25\%, 50\%, and 75\% connectivity, (2) random 3-regular graphs, and (3), complete graphs. Generating five instances per vertex-count and graph type (considering each {Erd\H{o}s-R\'{e}nyi}-connectivity as its own type), this results in a dataset of 444 graphs.

As baselines, we employ the current state-of-the-art HPC QAOA simulator QOKit as well as standard QAOA implementations in Qiskit and Pennylane. For Qiskit and Pennylane, cuStateVec is used to run the experiments on a GPU. The circuit simulation for QOKit is based on numba (which translates Python code into machine code upon compilation using Just-in-Time compilation and can natively be run on GPUs~\cite{10.1145/2833157.2833162}), as the cuStateVec variant of QOKit is not implemented in the currently available version of their code. While this limits the comparability of our results to the results published for QOKit, our methodology shows that all of our modifications to the QAOA simulation yield theory-proven improvements over QOKit, even when cuStateVec was executable for QOKit.

All experiments are run on a high-end consumer-grade system running EndeavourOS Linux x86\_64 with Linux Kernel version 6.8.7-arch1-1, 64GB of RAM, an AMD Ryzen 7 3700X CPU (16 cores @ 3.600 GHz), and an NVIDIA GeForce RTX 3090 GPU. All executions are started from within a Python script.
 
\subsection{Runtime of a single QAOA circuit simulation} \label{subsec:runtimeplainqaoa}
To examine the runtime of a plain QAOA circuit execution closing with a measurement of the expectation value, we compare CUAOA with all three baselines (QOKit, Qiskit, and Pennylane) in \Cref{fig:eval:results:expval:depth6}. For Pennylane, a memory allocation error occurred for problem instances exceeding 26 vertices, resulting in only 391 graphs being run successfully. Aside from this technical detail, we can observe that CUAOA performs best for all runs, even outperforming the state-of-the-art baselines by orders of magnitude in all small to medium-sized problem instances. It becomes evident that the effect of exponential runtime scaling only starts to manifest at around 16 qubits for CUAOA.

In-line with theoretical considerations, the application of the mixer unitary is eventually the biggest bottleneck for the runtime, as can be seen by comparing \Cref{fig:eval:results:expval:depth6} with the results from sampling-based QAOA runs displayed in \Cref{fig:eval:results:sample:depth6}, where the slope of the runtimes of all simulators become identical for problems above 20 qubits. The main reason that prevents further speedups is the sequential application of all mixer gates. Since every single mixer gate application modifies the memory of the entire statevector, parallelization is impossible.

Finally, the fact that the runtime of CUAOA is up to an order of magnitude better than QOKit's for problem instances below 20 qubits is necessarily the consequence of our CUDA-native implementation, as well as the optimized computations we introduced in \Cref{sec:methodology}. For problem instances beyond 20 qubits, this reduced overhead apparently marginalizes, leading to roughly equal runtimes, but with CUAOA still outperforming QOKit.

\begin{figure}[htbp]
    \centering
    \includegraphics[width=\columnwidth]{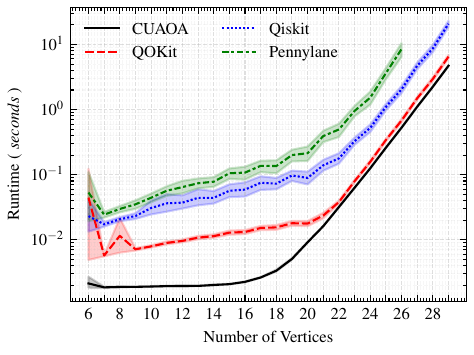}
    \caption{Runtime wrt. expectation value for QAOA with $p = 6$.}
    \label{fig:eval:results:expval:depth6}
\end{figure}

To evaluate the runtime of sampling bitstrings from the resulting statevector of the QAOA, \Cref{fig:eval:results:sample:depth6} displays the results for a QAOA run with $p=6$ and 1024 shots. To allow for better comparison to QOKit, whose available implementation at the time this article is published does not support sampling, we implemented a minimal-effort Python script. For this, we utilize QOKit's functionality to extract the probabilities of the statevector to the CPU. Subsequently, we use the standard Python random number generator to sample from the array containing the associated cumulative probabilities (which was computed using the \texttt{cumsum} function of numpy).

The results of CUAOA mirror those of the expectation value, showing the high degree of efficiency of both, i.e., not exceeding the runtime of the mixer application. While Qiskit performs quite well, Pennylane is significantly worse compared to the runtimes of the expectation value. The reasons for this are somewhat unclear, but indicate different implementations for sampling, especially for smaller circuit sizes. As expected, our CPU-based implementation of sampling for QOKit is hardly competitive as soon as the dimensionality of the statevector increases.

\begin{figure}[htbp]
    \centering
    \includegraphics[width=\columnwidth]{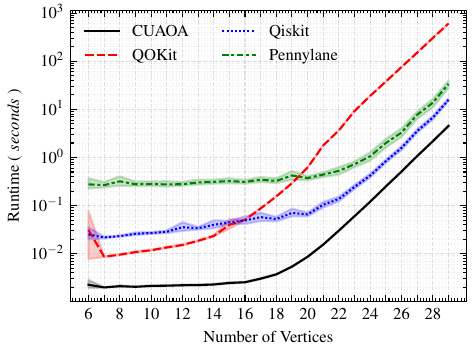}
    \caption{Runtime for sampling with 1024 shots and $p = 6$.}
    \label{fig:eval:results:sample:depth6}
\end{figure}

\subsection{Parameter Training}
Examining gradient-based parameter training, we now study the runtime of gradient computation for all parameters. As neither QOKit nor the cuStateVec-based Qiskit implementation allows for a computation of the gradient in the form of their available implementation at the time this paper is published, our experiment is restricted to a comparison of Pennylane and CUAOA. As the Pennylane implementation also implements the adjoint method, this is a fair comparison to our QAOA-aware enhanced version of the adjoint method. Analog to earlier evaluation runs, Pennylane again fails to execute graphs beyond 26 vertices, resulting in only 391 successfully executed problem instances. \Cref{fig:eval:results:grads:depth6} clearly shows that CUAOA outperforms Pennylane by multiple orders of magnitude for problem instances up to 18 qubits, being roughly 100 times faster in almost all runs. For a larger number of qubits, this gap closes to about a 10-fold speedup, with a significant runtime increase at around 18 qubits analog to what has manifested in the plain circuit evaluation in \Cref{subsec:runtimeplainqaoa}.

\begin{figure}[htbp]
    \centering
    \includegraphics[width=\columnwidth]{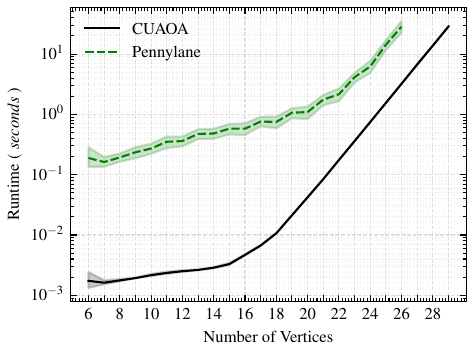}
    \caption{Runtime for gradient calculation using the adjoint differentiation method with $p = 6$.}
    \label{fig:eval:results:grads:depth6}
\end{figure}


Lastly, we also provide an implementation of the gradient based version of the optimizers L-BFGS (natively)~\cite{Liu1989} and BFGS (through scipy)~\cite{d70f72bb6d4749a5a8e29137299f32ed}. Additional experiments (not displayed here for brevity) show, that CUAOA is still up to two orders of magnitude faster than Pennylane when using the same optimizer (BFGS), essentially mirroring the results of \Cref{fig:eval:results:grads:depth6}.

\section{Conclusion}
\label{sec:conclusion}
In this paper, we proposed a classical high performance CUDA-based QAOA circuit simulator (CUAOA) oriented towards single-GPU usage. By exploiting speedups enabled through the diagonal structure of the cost operator at multiple stages of the QAOA simulation, i.e., (1) the computation and application of the cost operator, (2) the computation of the expectation value, and (3), providing a QAOA-specialized gradient computation method based on adjoint differentiation, our proposed implementation of the CUAOA outperformed the state-of-the-art QAOA simulator QOKit by an order of magnitude (i.e., a $10$-fold speedup) for small to medium sized problem instances. For large scale problem instances above 20 qubits, our approach also performed better than QOKit but equally suffers from the dominating runtime of the mixer operator. Notably, CUAOA offers significantly more functionality than QOKit for the key applications of (1) sampling from the statevector and (2) a GPU-based gradient computation. Further, our gradient computation runs about two orders of magnitude faster than the respective state-of-the-art approach. In conclusion, our approach can be regarded as the new state of the art for single-GPU QAOA simulation, as it outperformed all baselines up to multiple orders of magnitude in a representative evaluation.

In future work, our approach could be extended towards multi-GPU scenarios, which would require mostly additional implementation while relying on the same theoretical insights. Further, one could natively implement constraint-preserving mixers, by directly reducing the search space, which could significantly reduce numerical simulation runtime for heavily constrained problems.

\section*{Acknowledgment}
This paper was partially funded by the German Federal Ministry for Economic Affairs and Climate Action through the funding program "Quantum Computing -- Applications for the industry" based on the allowance "Development of digital technologies" (contract number: 01MQ22008A) and through the Munich Quantum Valley, which is supported by the Bavarian state government with funds from the Hightech Agenda Bayern Plus.

\bibliographystyle{IEEEtranDoi}  
\bibliography{bstcontrol,references} 

\end{document}